\documentclass[epj]{webofc}
\usepackage[utf8]{inputenc}
\usepackage[varg]{txfonts}   
\usepackage{booktabs}
\usepackage{xcolor}
\definecolor{darkred}{rgb}{0.4,0.0,0.0}
\definecolor{darkgreen}{rgb}{0.0,0.4,0.0}
\definecolor{darkblue}{rgb}{0.0,0.0,0.4}
\usepackage[bookmarks,linktocpage,colorlinks,
    linkcolor = darkred,
    urlcolor  = darkblue,
    citecolor = darkgreen]{hyperref}
\usepackage{subfigure}
\wocname{EPJ Web of Conferences}
\woctitle{Lattice2017}
\graphicspath{{./Figs/}}
\newcommand{\APF}{\langle{e^{i\theta}}\rangle_\mathrm{pq}}
\begin{document}
%
\selectlanguage{english}
\title{%
Path optimization method for the sign problem
\footnote{Report No.: YITP-17-128, KUNS-2709}
}
\author{%
\firstname{Akira} \lastname{Ohnishi}\inst{1}\fnsep\thanks{Speaker, \email{ohnishi@yukawa.kyoto-u.ac.jp}}
\and
\firstname{Yuto} \lastname{Mori}\inst{2} \and
\firstname{Kouji}  \lastname{Kashiwa}\inst{1}
}
\institute{%
Yukawa Institute for Theoretical Physics, Kyoto University,
Kyoto 606-8502, Japan
\and
Department of Physics, Faculty of Science, Kyoto University,
Kyoto 606-8502, Japan
}
\abstract{%
We propose a path optimization method (POM) to evade the sign problem
in the Monte-Carlo calculations for complex actions.
Among many approaches to the sign problem,
the Lefschetz-thimble path-integral method and the complex Langevin method
are promising and extensively discussed.
In these methods, real field variables are complexified
and the integration manifold is determined by the flow equations
or stochastically sampled.
When we have singular points of the action
or multiple critical points
near the original integral surface,
however, we have a risk to encounter the residual and global sign problems
or the singular drift term problem.
One of the ways to avoid the singular points is 
to optimize the integration path
which is designed not to hit the singular points of the Boltzmann weight. 
By specifying the one-dimensional integration-path
as $z=t+if(t) (f\in \mathbb{R})$
and by optimizing $f(t)$ to enhance the average phase factor,
we demonstrate that we can avoid the sign problem
in a one-variable toy model
for which the complex Langevin method is found to fail.
In this proceedings,
we propose POM and discuss how we can avoid the sign problem in a toy model.
We also discuss the possibility to utilize the neural network
to optimize the path.
}
\maketitle
\section{Introduction}\label{intro}

Solving the sign problem for complex actions is one of the grand challenges
in quantum many-body theories.
It is the largest obstacle to explore the phase diagram 
at finite densities from first principles.
Since the lattice QCD at finite baryon density has the sign problem,
we cannot obtain precise predictions on dense matter from lattice QCD.
As a result, we do not yet know the location of the QCD critical point
as well as the critical density to the quark matter at high density.
Even the order of the phase transition at high density is not known.
This is not only a theoretical problem, but also a phenomenologically
important question. The existence of the first order phase transition
at high density generally induces the softening of the equation of state,
which may be detected
in heavy-ion collisions via collective flows~\cite{%
Adamczyk:2014ipa,
Nara:2016phs,
Nara:2017qcg
}
or conserved charge cumulants~\cite{%
Adamczyk:2013dal
},
or in the hypermassive neutron star properties which would be
observed in binary neutron star mergers~\cite{%
TheLIGOScientific:2017qsa
}.

In order to attack the sign problem,
there have been many attempts such as
the Taylor expansion around zero density~\cite{Allton:2005gk},
the analytic continuation from the imaginary chemical potential~\cite{deForcrand:2002hgr,DElia:2002tig},
the canonical method based on calculations at imaginary chemical potential~\cite{Ejiri:2008xt,Nakamura:2015jra}
and
the strong coupling expansion in the mean field treatment~\cite{%
Ilgenfritz:1984ff,
Nishida:2003fb,
Fukushima:2003vi,
Kawamoto:2005mq,
Miura:2009nu,
Nakano:2009bf,
Nakano:2010bg,
Miura:2016kmd
}
or with the Monte-Carlo calculation~\cite{
Karsch:1988zx,
deForcrand:2009dh,
deForcrand:2014tha,
Ichihara:2014ova,
Ichihara:2015kba
}.
Recent developments in the sign problem include
the complex Langevin method (CLM)~\cite{%
Parisi:1980ys,%
Parisi:1984cs,%
Aarts:2009uq
},
the Lefschetz thimble method (LTM)~\cite{%
Witten:2010cx,%
Cristoforetti:2012su,%
Fujii:2013srak
},
and the generalized Lefschetz thimble method (GLTM)~\cite{%
Fukuma:2017fjq,%
Alexandru:2017oyw
}.
These methods are based on complexified field variables.
By integrating the Boltzmann weight $\exp(-S)$ ($S \in \mathbb{C}$)
on the shifted path from the original real axis,
we can suppress the cancellation
coming from the rapidly oscillating complex phase of the Boltzmann weight.
In LTM, integral is performed over thimbles defined by the flow equation
for complexified variables. Since the imaginary part of the action
is constant on one thimble, a large part of the sign problem is removed.
It should be also noted that the LTM is based on
the Cauchy(-Poincare) theorem, 
which states that integral of an analytic (holomorphic) function is independent
of the integral path as long as it is shifted from the original path
without going across singular points such as poles.
Still, we have problems in LTM.
The integration measure (Jacobian) can contain the complex phase (residual
sign problem),
and contributions from different thimbles
can kill the partition function
(global sign problem).
In addition, the flow equation blows up somewhere~\cite{Tanizaki:2017yow},
then it is not easy to perform full integration over thimbles.
Because of these reasons, LTM has not yet been applied to finite density QCD.
%
By comparison, CLM is a powerful tool and has been applied also to QCD.
In CLM, one solves the complex Langevin equation for complexified variables.
The fictitious time average is proven to agree with the exact results
in many (lucky) cases.
Yet, there have been many problems in CLM.
First, the evolution by the complex Langevin equation can easily enter
the region far from the original real axis (excursion problem),
since there is no "minimum" in analytic functions in the complex variable space.
The excursion problem may be cured by the gauge cooling method~\cite{%
Seiler:2012wz
}.
Second, converged results in CLM are not necessarily correct.
This problem is known to be caused by large drift term~\cite{%
Nagata:2016vkn
}.
We are interested in the QCD {\em phase} diagram at finite density,
the phase transition needs to be addressed,
and we cannot avoid to perform integration around the singular points.
For example, there are many singular points close to the real axis
in the complexified chiral field
even in a mean field treatment of the Nambu-Jona-Lasinio model~\cite{%
Mori:2017zyl
}.
Is there any way to obtain the integral path without solving the flow equation
and without suffering from singular points ?

One of the possible ways would be to optimize the path variationally.
We first prepare the parameterized path appropriately (trial function).
The path is optimized to minimize the function (cost function),
which reflects the seriousness of the sign problem.
We refer to this method,
the {\em path optimization method (POM)}~\cite{%
Mori:2017pne,
Mori:2017nwj
}.
Now the sign problem is converted into an optimization problem,
in which various methods have been developed.

In this proceedings, we introduce POM and demonstrate it
in a one-variable toy model having a serious sign problem~\cite{Mori:2017pne}.
In Sec.~\ref{Sec:POM}, we explain the basic idea of POM,
and introduce the trial function, cost function, and optimization.
In Sec.~\ref{Sec:Results}, we apply POM
to a toy model introduced in Ref.~\cite{Nishimura:2015pba}.
Section \ref{Sec:Summary} is devoted to summary.

\section{Path optimization method}\label{Sec:POM}

In quantum statistics, the partition function $\mathcal{Z}$ is defined
as the integral of the Boltzmann weight $e^{-S}$
over all the integration variables,
represented by $x \in \mathbb{R}$.
For a complex action $S \in \mathbb{C}$,
the partition function is less than the phase quenched one,
and its ratio is referred to as the average phase factor, $\APF$.
\begin{align}
\mathcal{Z}
=\int_{\mathcal{C}_{\mathbb{R}}} Dx\, \exp(-S[x])
\ ,
\quad
\langle{e^{i\theta}}\rangle_\mathrm{pq}
= \frac{\int Dx\, e^{-S}}{\int Dx\, |e^{-S}|}
= \frac{\int Dx\, |e^{-S}|\,e^{i\theta}}{\int Dx\, |e^{-S}|}
\ .
\end{align}
When the imaginary part of the action is a rapidly oscillating
function of $x$, the Boltzmann weight cancels with each other
and the average phase factor becomes small, $|\APF| \ll 1$.
The cancellation becomes more serious with increasing degrees of freedom, $N_D$.
We need to invoke the Monte-Carlo (MC) technique to calculate observables
with large $N_D$,
and the MC results always contain errors proportional to $1/\sqrt{N_\text{MC}}$
with $N_\text{MC}$ being the number of MC samples.
Since the average phase factor exponentially decreases with increasing $N_D$, 
we need exponentially large number of MC samples.
This is the sign problem.

\begin{figure}
\begin{center}
\includegraphics[width=12cm,clip,bb=35 20 760 220]{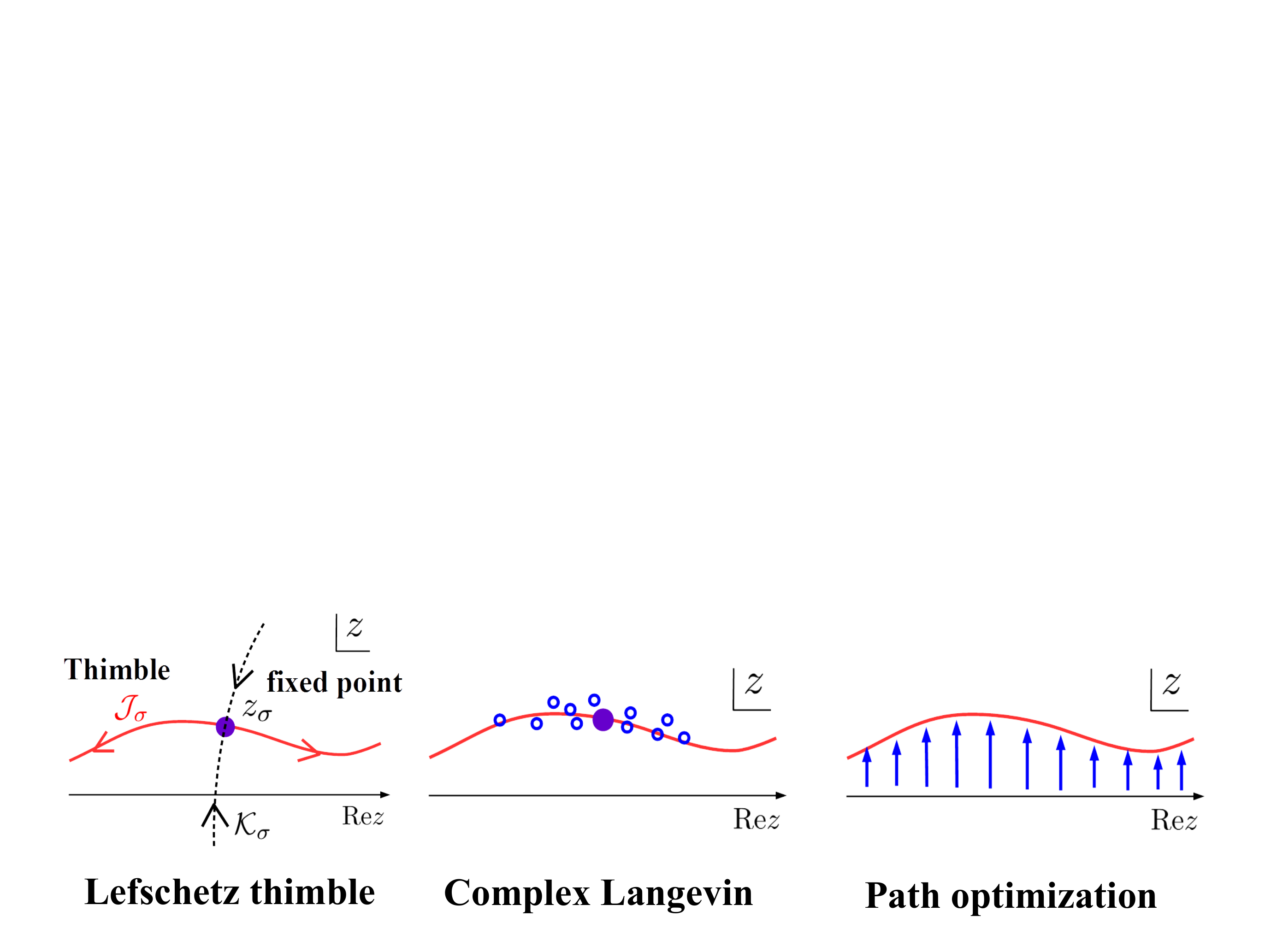}
\end{center}
\caption{Integral path and distribution of samples in LTM, CLM and POM.}
\label{Fig:path}
\end{figure}

Provided that the action is an analytic function of the integration variables, 
it is possible to complexify the integration variables
and to shift the integral path $\mathcal{C}$
off the real axis $\mathcal{C}_\mathbb{R}$,
\begin{align}
\mathcal{Z}
=\int_\mathcal{C} Dz\, \exp(-S[z])
=\int_{\mathcal{C}_\mathbb{R}} Dt\, J(z(t)) \exp(-S[z(t)])
 \ .
\end{align}
In the most right-hand side, we have rewritten $Dz=Dt\,J$
where $t\in \mathbb{R}$ and $J$ is the Jacobian.
The path shift should not go across singular points
of the Boltzmann factor $e^{-S}$,
and we assume that the integral at $\mathrm{Re}\, z \to \pm \infty$
are negligible.
Under this condition, the Cauchy(-Poincare) theorem 
tells us that the partition function is independent of the integral path.
By comparison, the phase quenched partition function depends on the path.
Thus our task to evade the sign problem is to find the integral path
which provide the large enough average phase factor.

In the path optimization method (POM)~\cite{%
Mori:2017pne,
Mori:2017nwj
},
we optimize the integral path
variationally so as to enhance the average phase factor.
%
First, we prepare the trial function, which parameterize the integral path. 
In the one variable case, for example, we can expand the $z(t)$
by a complete set $\{H_n\}$,
\begin{align}
z(t)=x(t)+iy(t)=t + \sum_n (c_n^{(x)}+ic_n^{(y)})\,H_n(t)\ .
\end{align}
Next, we define the cost function,
which should reflect the seriousness of the sign problem.
Here we use the following cost function,
\begin{align}
F[z(t)]=&\frac12 \int dt \left|e^{i\theta(t)}-e^{i\theta_0}\right|^2
\left| J(t)\,e^{-S[z(t)]} \right|
= \left|\mathcal{Z}\right|\left[ \left|\APF\right|^{-1} - 1 \right]
\ ,
\end{align}
where $\theta(t)$ and $\theta_0$ are the complex phase
of $J\,e^{-S}$ and $\mathcal{Z}$, respectively.
Since the partition function is independent of the path,
reducing $F$ corresponds to enhancing the average phase factor.
Finally, we optimize the integral path from the original one
by tuning the parameters, $c_n^{(x)}$ and $c_n^{(y)}$,
as schematically shown in Fig.~\ref{Fig:path}.
There have been many optimization methods developed so far.
For simple problems, we can apply the steepest descent method,
and for complicated systems with large degrees of freedom,
machine learning technique would be promising.

There are two comments in order.
One of them is the comparison with other methods.
Compared with LTM, we start from the original integral path
and it is not necessary to find the fixed point
($\partial S/\partial z=0$) in advance in POM, as well as in GLTM.
Compared with CLM, we can avoid singular points,
by integrating along the optimized path,
provided that the path does not hit the singular points.
%
%
Another point is on the degree of optimization.
It should be noted that we can obtain the expectation values of observables
safely and precisely as long as the average phase factor is
clearly different from zero,
and it is not necessary to fully maximize the average phase factor.
We do not need to be fundamentalists
by trying to fully maximize the average phase factor.

\section{Application to a toy model}\label{Sec:Results}

Now we examine the validity and usefulness of POM
in a one-variable toy model proposed in Ref.~\cite{Nishimura:2015pba}.
The partition function is given by the one dimensional integral,
\begin{align}
\mathcal{Z}=\int dx (x+i\alpha)^p \exp(-x^2/2)
=\int dx\, e^{-S}
\ ,\quad
S(x)=x^2/2-p\log(x+i\alpha)
\ ,
\end{align}
where $p$ is a positive integer.
For large $p$ and small $\alpha$,
CLM fails to describe the exact results~\cite{Nishimura:2015pba}.
For large $p$, the complex phase oscillates very rapidly.
For small $\alpha$, the singular point of the action
($z=-i\alpha$) is close to the real axis.
Then the partition function frequently becomes zero
in the small $\alpha$ region, $\alpha<13$.
When $p$ is a positive integer,
the Boltzmann weight is zero and smooth at $z=-i\alpha$,
then it is not necessary to care in POM.
The relevant singular point, the singular point of the Boltzmann weight,
exists at $|\mathrm{Im}z| \to \pm \infty$.
Nevertheless, the point $z=-i\alpha$ gives rise to a problem in CLM,
and it is referred to as the "singular points" in the later discussions.

We use here a simple trial function,
\begin{align}
z(t)=t+i\left[c_1 \exp(-c_2^2 t^2/2)+c_3\right]
\label{Eq:trial}
\ ,
\end{align}
and optimize the path by the steepest descent method,
$dc_i/dt=-\partial F/\partial c_i$.
The integration is performed using the double exponential formula.

\begin{figure}
\begin{center}
\includegraphics[width=7cm]{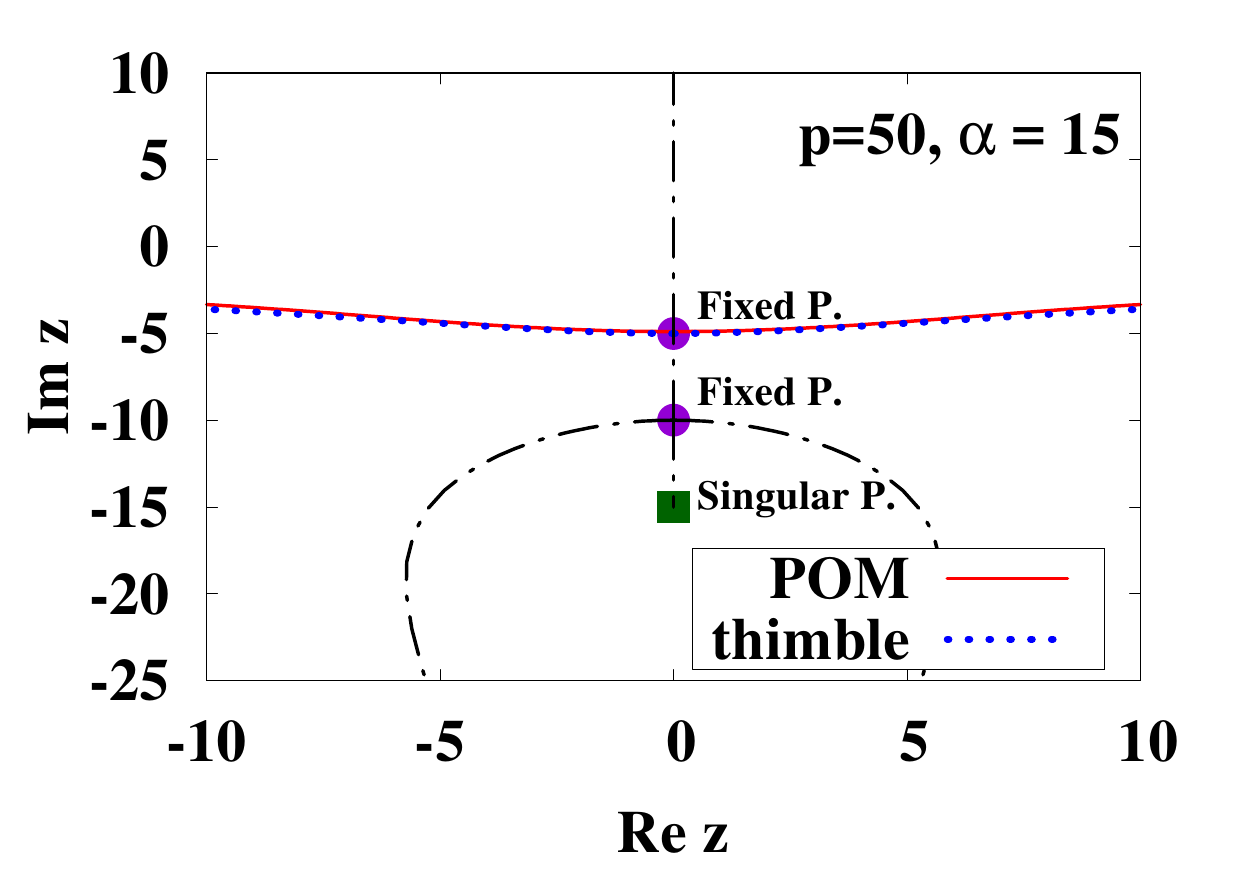}%
\includegraphics[width=7cm]{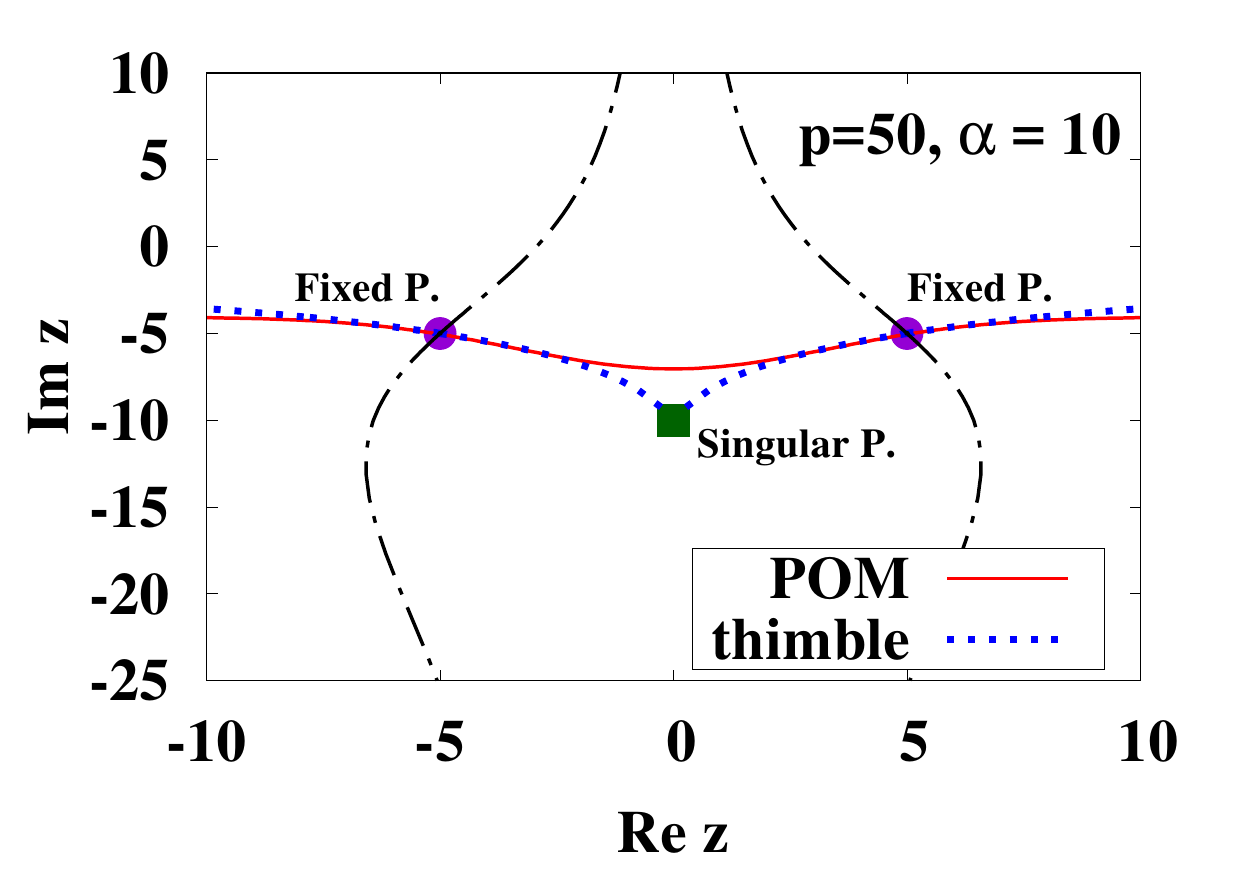}
\end{center}
\caption{Optimized integral path in POM and the Lefschetz thimbles
in a toy model for $p= 50$, $\alpha=15$ (left) and $\alpha=10$ (right).
Filled circle (square) shows the fixed (singular) point.
Dot-dashed lines are steepest ascent paths.}
\label{Fig:opt}
\end{figure}

In Fig.~\ref{Fig:opt}, we show the optimized path
at $\alpha=15$ and $\alpha=10$ in comparison with thimbles.
In both cases, we take $p=50$.
We find that the optimized path is close to the thimble(s),
especially at around the fixed points,
and it avoids the singular point.
At $\alpha=15$ where the sign problem is less serious,
there is only one relevant thimble which is off the singular point.
At $\alpha=10$, both of two thimbles are relevant,
and they terminate at the singular point.
The optimized path goes through the two fixed points,
but deviates from the singular point.
While we did not require,
it is natural for the optimized path to go through the fixed points,
around which the phase oscillation is mild.
It is also natural that the optimized path does not come close
to the "singular point" of the action, where the statistical weight is small.

\begin{figure}
\begin{center}
\includegraphics[width=7cm]{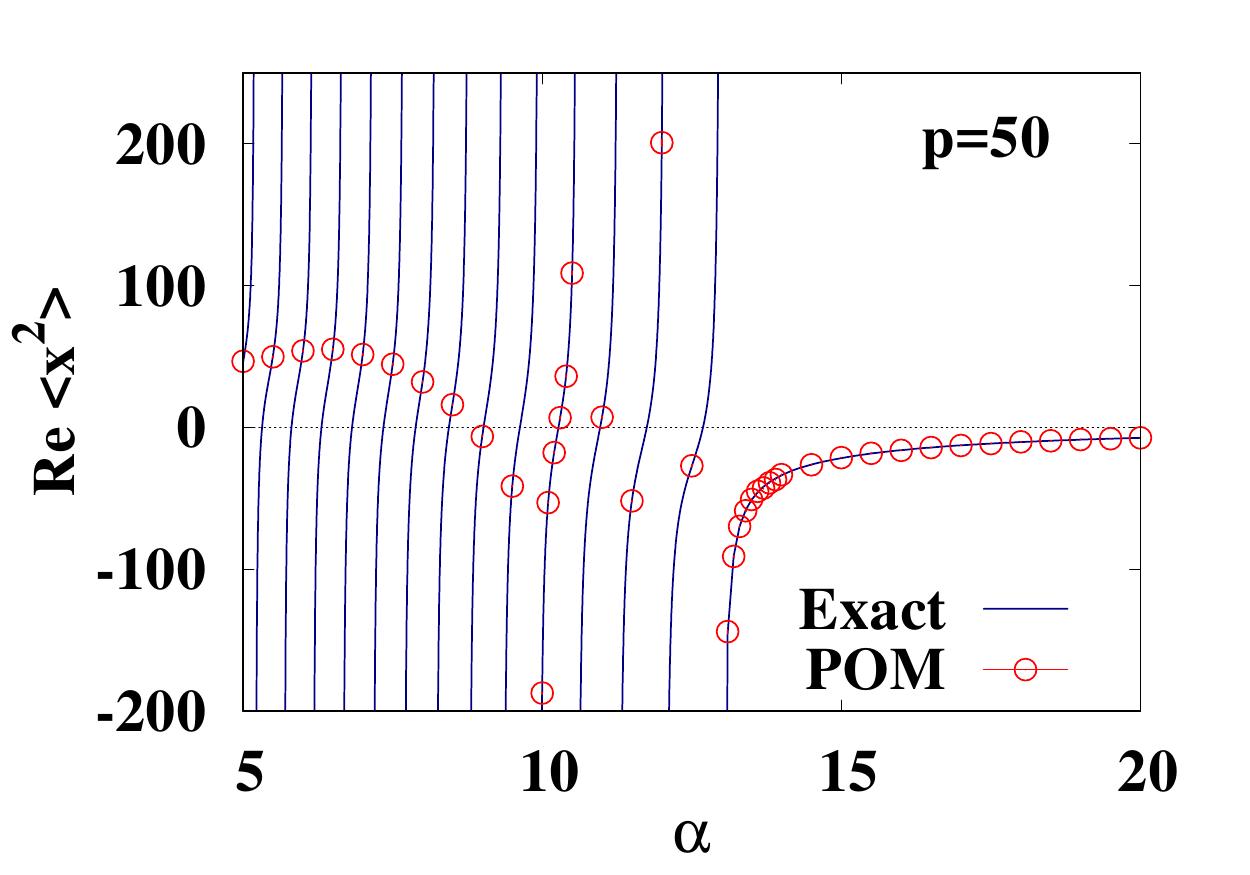}%
\includegraphics[width=7cm]{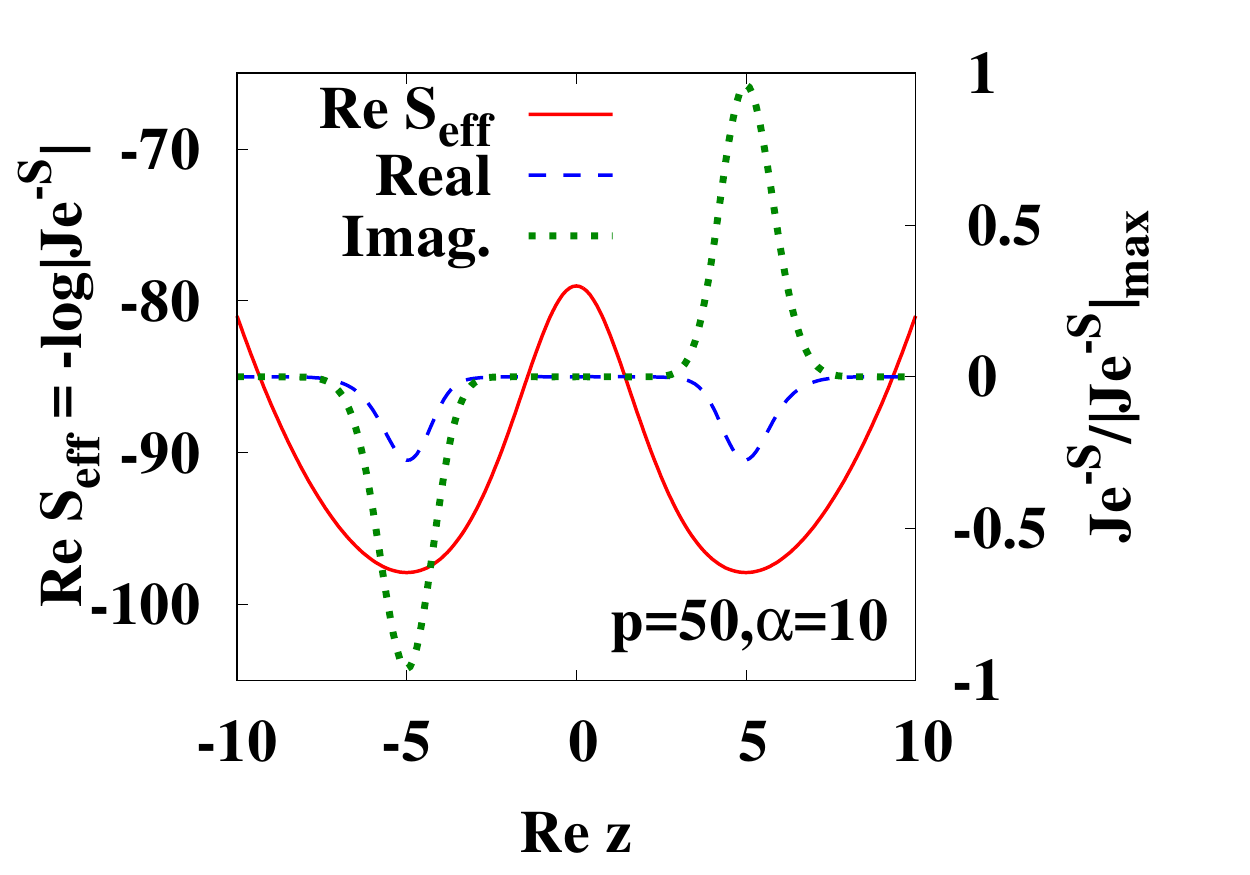}
\end{center}
\caption{
Left: Expectation values of $\mathrm{Re}\,x^2$ for $p = 50$
as a function of $\alpha$.
Symbols show results obtained in the hybrid Monte-Carlo method
on the optimized path in POM.
Dotted lines show exact results.
Errors are estimated in the Jackknife method.
Right: Statistical weight $J\,e^{-S}$ on the optimized path.
Solid line shows the real part of the effective action defined as,
$\mathrm{Re}\,S_\mathrm{eff}=-\log|J\,e^{-S}|$.
Dashed and dotted lines show the real and imaginary part
of $J\,e^{-S}$ normalized by the maximum value of $|J\,e^{-S}|$.
}
\label{Fig:x2}
\end{figure}

Next, we show the expectation value of $\text{Re}\,x^2$
in the left panel of Fig.~\ref{Fig:x2}.
The expectation value of an observable $\mathcal{O}$ is also independent
of the path as long as it is an analytic function,
\begin{align}
\langle\mathcal{O}\rangle
=\frac{1}{\mathcal{Z}}\int_\mathcal{C} Dz\, \mathcal{O}(z)\,e^{-S[z]}
\ .
\end{align}
We have used the hybrid Monte-Carlo method on the optimized path
to calculate the expectation value
in order to demonstrate the usefulness of POM in practical calculations.
The obtained results well agree with the exact ones
even in the severe region of the sign problem.
For example,
we can explain the exact results around $\alpha=10$.
At $\alpha \simeq 9.95$ and $\alpha \simeq 10.59$,
the partition function vanishes and the expectation value diverges.
Errors evaluated by using the Jackknife method are smaller
than the symbol size, as you can find by magnifying the figure.

We should confess here that this high precision is achieved 
in part by the sampling method using symmetry.
In the hybrid MC sampling,
$\pm t$ points are taken at the same time
based on the reflection symmetry of the phase quenched
statistical weight,
$W(t) = W(-t)$, where $W(t)=|J(t)e^{-S(t)}|$.
In the right panel of Fig.~\ref{Fig:x2},
we show the real part of the effective action,
$\mathrm{Re}\,S_\mathrm{eff}\equiv-\log|J\,e^{-S}|=-\log W(t)$,
on the optimized path as a function of $t$ at $\alpha=10$.
Compared with the minimum value (at the fixed points),
the effective action at the barrier is higher by around 20.
We also show the real and imaginary parts of $J\,e^{-S}$
normalized by the maximum value of $|J\,e^{-S}|$.
The statistical weight is dominated by the imaginary part,
which takes positive and negative values
around the fixed points in the region of $t=\mathrm{Re}\,z>0$ and $t<0$,
respectively, and cancel with each other in the integral.
The potential barrier between the two fixed points is so high
that the MC configuration around one fixed point cannot reach
the region around the other fixed point.
As a result,
the cancellation is forgotten
and the absolute value of the partition function is overestimated
without the simultaneous sampling,
and the expectation value of $x^2$ is underestimated.
Similar treatment is applied in GLTM~\cite{%
Fukuma:2017fjq,%
Alexandru:2017oyw
}.
It should be noted that the partition function can be zero
in POM even with the sampling using symmetry and also in the exact results.
We cannot (and should not) solve the global sign problem in POM.
In the case when there are two or more local minima separated by high barriers
and we do not know symmetry among the local minima,
we need to invoke the exchange MC~\cite{hukushima1996exchange}
technique carefully.
As asked at the conference,
if there are too many local minima contributing to the partition function
differently, it will be difficult to perform MC integrals.

\begin{figure}
\begin{center}
\includegraphics[width=12cm,clip,bb=50 50 710 540]{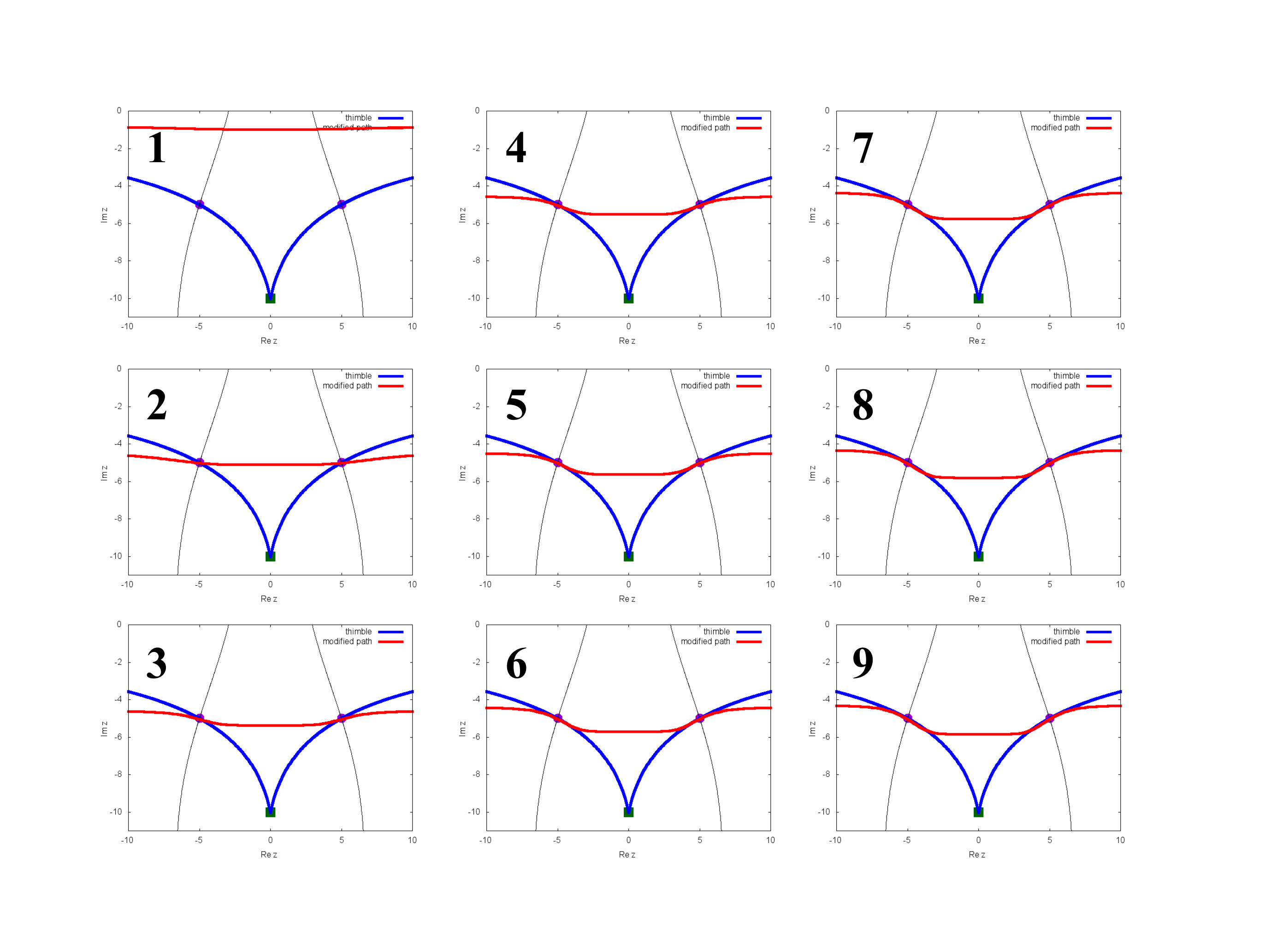}
\end{center}
\caption{Path optimization using a neural network.
Red curves show the paths during optimization,
and blue curves show Lefschetz thimbles.
Each panel shows the results during the optimization,
and the numbers at the top-left corner of the panel show the update steps
with some interval.
}
\label{Fig:NN}
\end{figure}

Readers may suspect that POM works only in simple systems
for which we can prepare an appropriate trial function.
We would like to make an objection to this suspicion.
In Fig.~\ref{Fig:NN}, we show the results of path optimization
at $\alpha=10$ using the neural network\footnote{
We have explained how we use the neural network in POM
in Ref.~\cite{Mori:2017nwj}. 
Interested readers are referred
to Ref.~\cite{Mori:2017nwj} and references therein.
}.
One of the merits to use the neural network is that we do not need to prepare
the form of the trial function.
The output $z(t)$ is given by the combination of linear transformation
and the activation function (such as the hyperbolic tangent),
then one can obtain a wide class of functions.
In the present calculation,
optimization starts from the original path.
The path first moves in the negative imaginary direction
and catches the fixed points,
and later bends to find the slope at which $\mathrm{Im}\,J\,e^{-S}$
is constant around the fixed points.
At around the fixed points,
optimized paths by the two methods agree with each other.
In other regions, two paths deviate from each other.
Since the partition function is dominated by the fixed point regions,
the above deviation leads to very small differences
in the average phase factor and the expectation values of observables.

\section{Summary}\label{Sec:Summary}

We have proposed a path optimization method (POM) to attack the sign problem.
We parameterize the integration path by the trial function,
the seriousness of the sign problem is represented by the cost function,
and the path is optimized to reduce the cost function.
If we can enhance the average phase factor
to a value clearly different from zero by the optimization,
it becomes possible to obtain the expectation value of any observable
safely and precisely.
In this way, the sign problem is regarded as the optimization problem.

We have examined POM in a one-variable toy model,
for which the complex Langevin model fails to give precise results.
The optimized path is found to agree with the Lefschetz thimble
around the fixed points, where the action is stationary.
The expectation value of an observable ($x^2$) calculated on the optimized path
agrees with the exact results even in the severe region of the sign problem.
Optimization can be performed by using a simply parameterized trial function
or by the neural network.

Application of POM to other actions with larger degrees of freedom is desired.
After the conference, we have applied POM with use of the neural network
as the optimization method to the $\phi^4$ theory~\cite{%
Mori:2017nwj
},
and have found that POM works efficiently.
We are also working on several other actions. Please stay tuned.

This work is supported in part by the Grants-in-Aid for Scientific Research
from JSPS (Nos. 15K05079, 15H03663, 16K05350),
the Grants-in-Aid for Scientific Research on Innovative Areas from MEXT
(Nos. 24105001, 24105008),
and by the Yukawa International Program for Quark-hadron Sciences (YIPQS).

%


\end{document}